\documentclass[a4paper,11pt]{article}
\usepackage{jinstpub} 
\DeclareMathOperator\erf{erf} 
\usepackage{lineno}

\title{\boldmath Skipper-CCD Sensors for the Oscura Experiment: Requirements and Preliminary Tests}

\author[1]{Brenda A. Cervantes-Vergara}
\author[15]{Santiago Perez}
\author[4]{Juan Estrada}
\author[4]{Ana Botti}
\author[4,8]{Claudio R. Chavez}
\author[8]{Fernando Chierchie}
\author[4]{Nathan Saffold}
\author[1]{Alexis Aguilar-Arevalo}
\author[2]{Fabricio Alcalde-Bessia}
\author[2]{Nicol\'as Avalos}
\author[3]{Oscar Baez}
\author[4]{Daniel Baxter}
\author[2]{Xavier Bertou}
\author[5]{Carla Bonifazi}
\author[4]{Gustavo Cancelo}
\author[6]{Nuria Castelló-Mor}
\author[7]{Alvaro E. Chavarria}
\author[3]{Juan Manuel De Egea}
\author[1]{Juan Carlos D'Olivo}
\author[9]{Cyrus Dreyer}
\author[4,10]{Alex Drlica-Wagner}
\author[9]{Rouven Essig}
\author[2]{Ezequiel Estrada}
\author[11]{Erez Etzion}
\author[12]{Paul Grylls}
\author[4]{Guillermo Fernandez-Moroni}
\author[9]{Marivi Fern\'andez-Serra}
\author[3]{Santiago Ferreyra}
\author[13]{Stephen Holland} 
\author[6]{Agustín Lantero Barreda}
\author[4]{Andrew Lathrop}
\author[12]{Ian Lawson}
\author[14]{Ben Loer}
\author[12]{Steffon Luoma}
\author[10,4]{Edgar Marrufo Villalpando}
\author[1]{Mauricio Martinez Montero}
\author[7]{Kellie McGuire}
\author[3]{Jorge Molina}
\author[10]{Sravan Munagavalasa}
\author[10]{Danielle Norcini}
\author[7]{Alexander Piers}
\author[10]{Paolo Privitera}
\author[15]{Dario Rodrigues}
\author[14]{Richard Saldanha}
\author[9]{Aman Singal}
\author[10]{Radomir Smida}
\author[16]{Miguel Sofo-Haro}
\author[3]{Diego Stalder}
\author[4]{Leandro Stefanazzi}
\author[4]{Javier Tiffenberg}
\author[7]{Michelangelo Traina}
\author[4]{Sho Uemura}
\author[17]{Pedro Ventura}
\author[6]{Rocío Vilar Cortabitarte}
\author[10]{Rachana Yajur}

\affiliation[1]{Universidad Nacional Aut\'onoma de M\'exico, Ciudad de M\'exico, M\'exico}
\affiliation[2]{Centro Atomico Bariloche, Rio Negro, Argentina}
\affiliation[3]{Facultad de Ingenier\'ia, Universidad Nacional de Asunci\'on, Paraguay}
\affiliation[4]{Fermi National Accelerator Laboratory, IL, USA}
\affiliation[5]{International Center of Advanced Studies and Instituto de Ciencias Físicas, ECyT-UNSAM and CONICET, Argentina}
\affiliation[6]{Instituto de Fisica de Cantabria, Santander, Spain }
\affiliation[7]{University of Washington, WA, USA}
\affiliation[8]{IIIE CONICET and DIEC Universidad Nacional del Sur, Argentina}
\affiliation[9]{Stony Brook University, NY, USA}
\affiliation[10]{University of Chicago, IL, USA}
\affiliation[11]{Tel Aviv University, Israel}
\affiliation[12]{SNOLAB, ON, Canada}
\affiliation[13]{Lawrence Berkeley National Laboratory, CA, USA}
\affiliation[14]{Pacific Northwest National Laboratory, WA, USA}
\affiliation[15]{Universidad de Buenos Aires, Buenos Aires, Argentina}
\affiliation[16]{Universidad Nacional de C\'ordoba, Instituto de F\'isica Enrique Gaviola (CONICET) and Reactor Nuclear RA0 (CNEA), C\'ordoba, Argentina.}
\affiliation[17]{Instituto de F\'isica, Universidade Federal do Rio de Janeiro, Rio de Janeiro, RJ, Brazil}

\emailAdd{brenda.cervantes@correo.nucleares.unam.mx}

\abstract{
Oscura is a proposed multi-kg skipper-CCD experiment designed for a dark matter (DM) direct detection search that will reach unprecedented sensitivity to sub-GeV DM-electron interactions with its 10~kg detector array. Oscura is planning to operate at SNOLAB with 2070~m overburden, and aims to reach a background goal of less than one event in each electron bin in the 2–10 electron ionization-signal region for the full 30 kg-year exposure, with a radiation background rate of 0.01 dru\footnote{1~dru (differential rate unit) corresponds to 1~event/kg/day/keV.}. In order to achieve this goal, Oscura must address each potential source of background events, including instrumental backgrounds. In this work, we discuss the main instrumental background sources and the strategy to control them, establishing a set of constraints on the sensors' performance parameters. We present results from the tests of the first fabricated Oscura prototype sensors, evaluate their performance in the context of the established constraints and estimate the Oscura instrumental background based on these results.
}

\begin{document}
\maketitle
\flushbottom

\section{The Oscura experiment} \label{sec:oscura}
Identifying the nature of dark matter (DM) is one of the most important missions of particle physics and astrophysics today, and direct-detection experiments play an essential role in this endeavor. The search for DM particles with masses up to a few orders of magnitude below the proton mass (``sub-GeV DM'') represents an important new experimental frontier that has been receiving increased attention, e.g.~\cite{Essig:2011nj, Essig:2013lka, Alexander:2016aln, Battaglieri:2017aum, BRNreport,Essig:2022dfa}. Typically, traditional direct-detection searches looking for DM particles scattering elastically off nuclei have very little sensitivity to sub-GeV DM \cite{Abdelhameed:2019hmk, Angloher:2017sxg, Petricca:2017zdp, Agnese:2015nto, Alkhatib:2020slm}. Improved sensitivity to DM masses well below the GeV scale is possible by searching for signals induced by inelastic processes~\cite{Essig:2011nj}. One of the most promising avenues is to search for one or a few ionization electrons that are produced by DM particles interactions with electrons in the detector~\cite{Essig:2011nj, BRNreport}.

Skipper-CCDs are among the most promising detector technologies for the construction of a large multi-kg experiment for probing electron recoils from sub-GeV DM. These ultra-low readout noise sensors, designed by the Lawrence Berkeley National Laboratory (LBNL) Microsystems Laboratory, allow for the precise measurement of the number of free electrons in each of the million pixels across the CCD~\cite{skipper2017}. This feature, combined with a low background rate, has allowed skipper-CCD experiments to set the strongest constraints to date within the direct dark matter searches on sub-GeV DM-electron interactions~\cite{sensei2018, sensei2019, SENSEI:2020dpa, DAMIC-M2023}, motivating the deployment of more massive detectors in the near future. Particularly, the SENSEI Collaboration has partially commissioned a $\sim$100~g skipper-CCD array at SNOLAB and the DAMIC-M Collaboration is aiming to build a $\sim$1~kg experiment at the Laboratoire Souterrain de Modane in the coming years.

\begin{figure}[h!]
    \centering
    \includegraphics[width=0.5\textwidth]{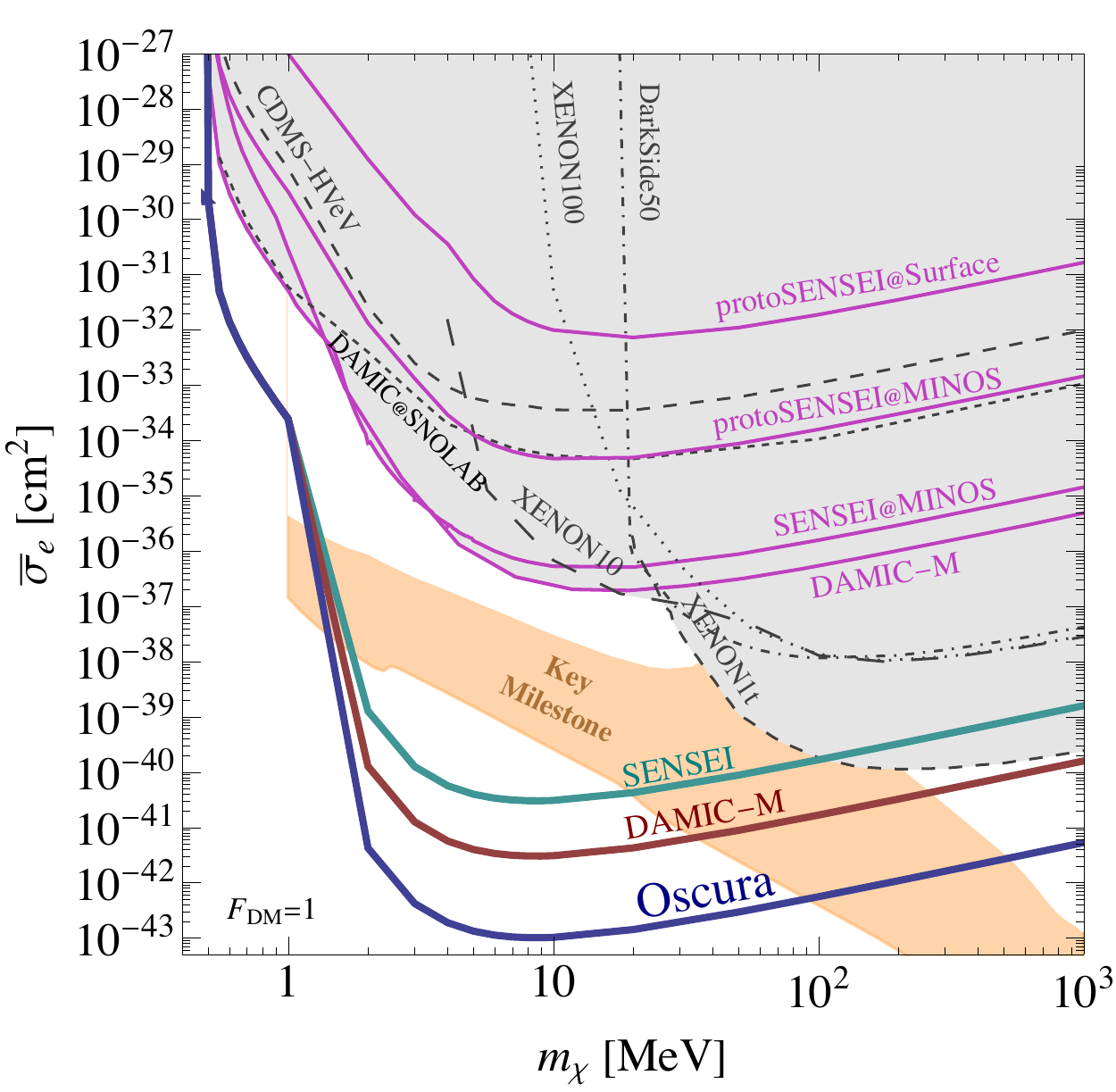}\hfill
    \includegraphics[width=0.5\textwidth]{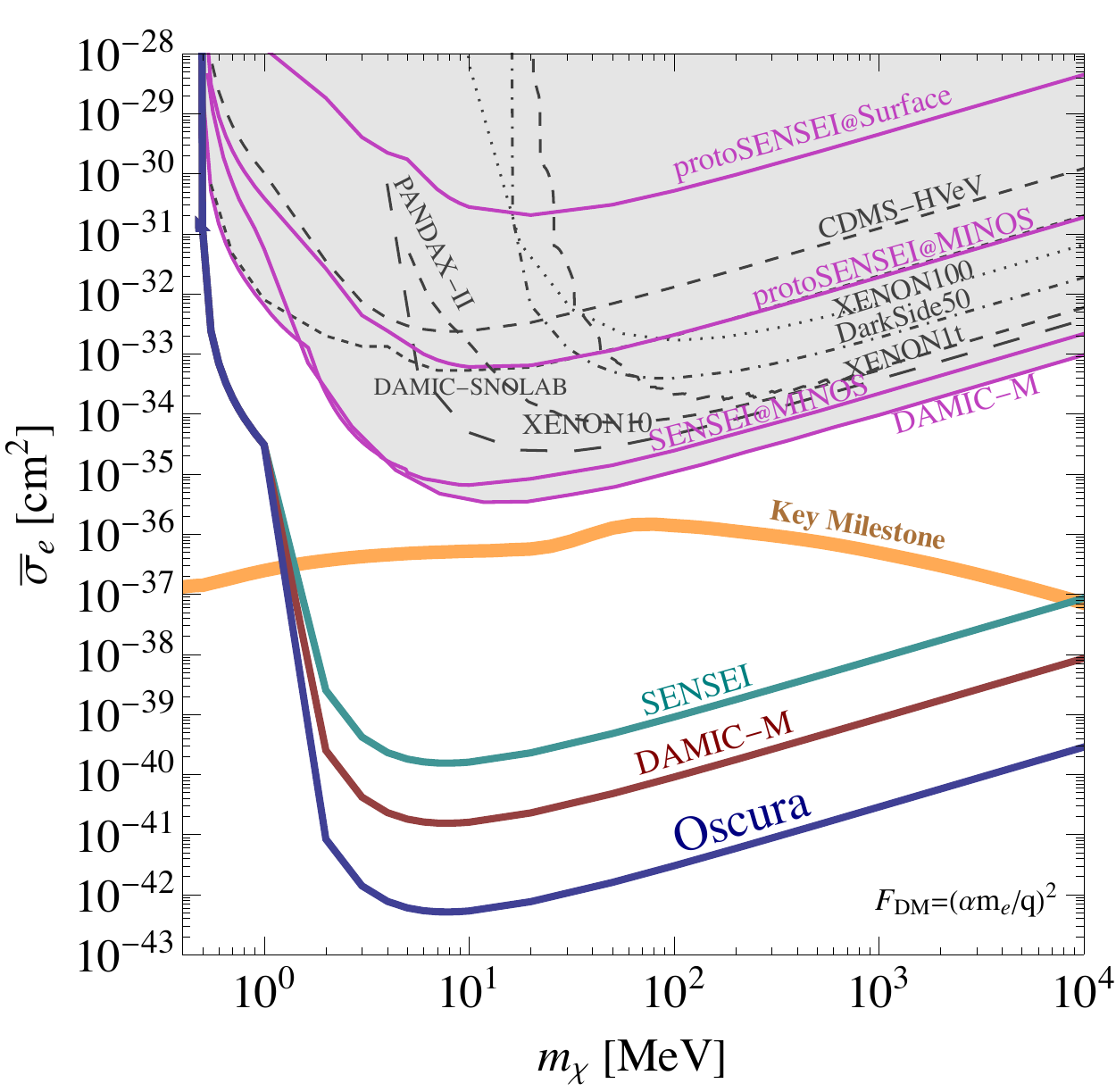}
    \caption{Approximate projected sensitivity for Oscura to DM-electron scattering at 90\% C.L. assuming a 30 kg-year exposure, zero background events with $2e^-$ or more, a $1e^-$ threshold and a fixed $1e^-$ event rate of $10^{-6} e^-$/pix/day (blue). To build this curve, 100\% efficiency was assumed for the reconstruction of events above $2e^-$. The left (right) plot assumes a heavy (light) mediator in the DM-electron interaction. Approximate projected sensitivities for SENSEI (DAMIC-M) are shown in cyan (red)~\cite{Essig:2011nj, Essig:2015cda, skipper2017, Settimo:2018qcm, Castello-Mor:2020jhd, BRNreport}. Existing constraints from skipper-CCDs from SENSEI~\cite{sensei2018, sensei2019, SENSEI:2020dpa} and DAMIC-M~\cite{DAMIC-M2023} are shaded in pink. Shaded gray regions are constrained by several other experiments (some shown explicitly)~\cite{Essig:2012yx, Essig:2017kqs, Angle:2011th, Aprile:2016wwo, Aprile:2019xxb, Agnes:2018oej, Agnese:2018col, Aguilar-Arevalo:2019wdi, Essig:2019xkx, Amaral:2020ryn, PandaX-II:2021nsg, XENON:2021myl}. Existing limits come directly from publications; reader should look at them for specific assumptions. Orange regions labeled “Key Milestone” represent well-motivated sub-GeV DM models, highlighted in the recommendations of the Basic Research Needs report~\cite{BRNreport}.
}
    \label{fig:projection-scattering}
\end{figure}

Oscura is a next-generation skipper-CCD DM search. It aims to collect a 30~kg-year exposure with less than one background event in each electron bin\footnote{Energy bin whose width is 3.745 eV, the mean ionization energy required for photons to produce an electron-hole pair in silicon~\cite{Ryan1973}.} in the 2–10 electron ionization-signal region; we will refer to this as the Oscura background \textit{goal}. To achieve it, a radiation background below 0.025 dru is needed, as well as a 1$e^-$ event rate below $1\times10^{-6} e^-$/pix/day coming from instrumental background sources (further discussed in Section~\ref{sec:instbkgs}). Oscura will probe unexplored regions in the parameter space of sub-GeV DM interacting with electrons. As an example, we show in Fig.~\ref{fig:projection-scattering} the approximate projected sensitivity for Oscura to DM-electron scattering through a ``heavy'' or ``ultralight'' mediator, particularly probing DM masses in the range of 500~keV to 1~GeV~\cite{Essig:2011nj, Graham:2012su, Essig:2015cda, Essig:2017kqs, Lee:2015qva}. For these projections, we assume the QeDark cross section calculation for DM-electron scattering~\cite{Essig:2015cda} and the astrophysical parameters considered in Ref.~\cite{SENSEI:2020dpa}.

We should emphasize that the Oscura experiment is building on existing efforts using skipper-CCDs to search for DM. All these experiments are developing the scientific and technical expertise to decrease the backgrounds. While we have set a stringent background \textit{goal}, we also consider the less stringent background \textit{requirement} of having less than one background event in each electron bin in the 3-10 electron ionization-signal region. A comment regarding Oscura science reach if unable to attain the background goal can be found in Section~\ref{sec:sumdisc}.

\subsection{Detector design} \label{sec:design}
The Oscura detector is a 10~kg silicon skipper-CCD array. To comply with standard fabrication processes, each sensor has 15~$\mu$m $\times$ 15~$\mu$m pixels and the standard thickness of 200~mm silicon wafers (725~$\mu$m). With these pixel dimensions, Oscura will need a 26 gigapixel array to achieve 10~kg of active mass.

The whole detector design and shielding are based on the constraints for reaching the Oscura background goal. The instrumental background restricts the sensors' performance parameters, and it will be deeply discussed in Section~\ref{sec:instbkgs}. For the radiation background, Oscura plans to reach 0.01 dru, which corresponds to a significant improvement over previous CCD experiments~\cite{DAMIC2016, DAMIC2020, sensei2019, 2020DAMICM}. This mandates strict control of all materials selected for the experiment and imposes a cosmogenic activation control requirement, particularly significant for the sensors (less than five days of sea level exposure equivalent after tritium removal~\cite{saldanha2020cosmogenic}).

\begin{figure}[h!] 
\centering
\includegraphics[width=0.25\linewidth]{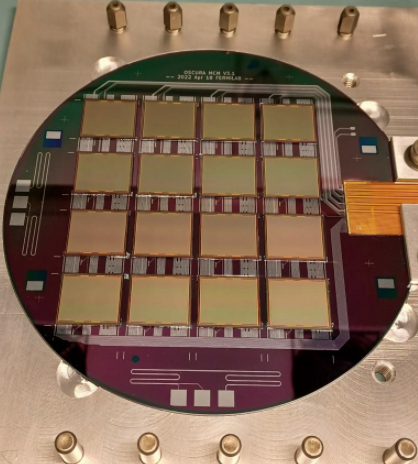} \hfill
\includegraphics[width=0.37\linewidth]{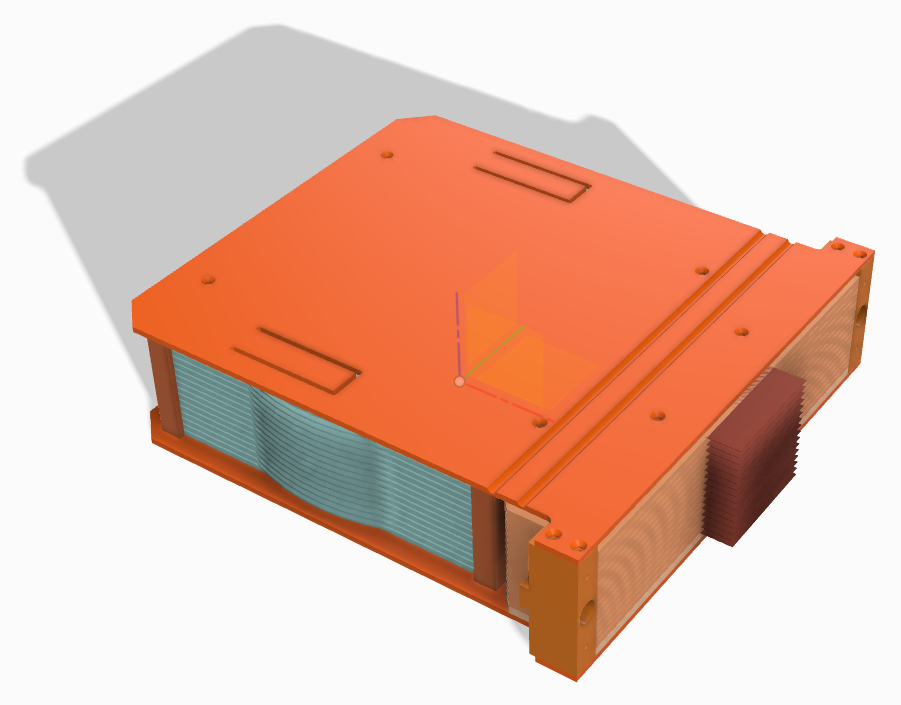} \hfill
\includegraphics[width=0.3\linewidth]{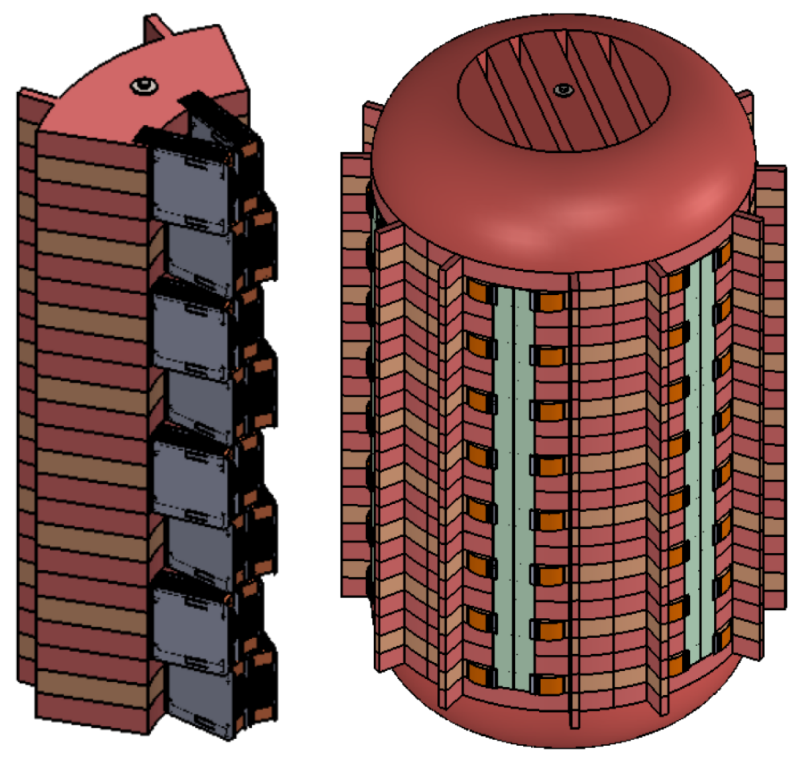}
\caption{Left) A fully assembled Si-MCM in a copper tray. Center) Oscura Super Module design with 16 MCMs supported and shielded with electroformed copper. Right) Model showing one of the columnar segments with 16 SMs each and the full assembly of all six segments to form the full cylindrical Oscura detector payload.}
\label{fig:MCM}
\end{figure}

The Oscura design is based on 1.35 Mpix sensors ($1278\times1058$ pixels) packaged on a Multi-Chip-Module (MCM), see Fig.~\ref{fig:MCM} (left). Each MCM consists of 16 sensors epoxied to a 150~mm diameter silicon wafer, with traces connecting the sensors to a low-radiation background flex cable~\cite{arnquist2020ultra, Arnquist:2023gtq}. MCMs will be integrated into Super Modules (SMs), where each SM will hold 16 MCMs using a support and shielding structure of custom ultrapure electro-deposited copper~\cite{electroformedcopper}, see Fig.~\ref{fig:MCM} (center). The Oscura experiment needs $\sim$80~SMs to reach 10~kg of active mass. The full detector payload consists of 96 SMs, assuming a yield above 80\%, surrounded by an internal copper and lead shield, arranged in six columnar slices forming a cylinder, see Fig.~\ref{fig:MCM} (right).

\begin{figure}[h!]
    \centering
    \includegraphics[width=0.48\textwidth]{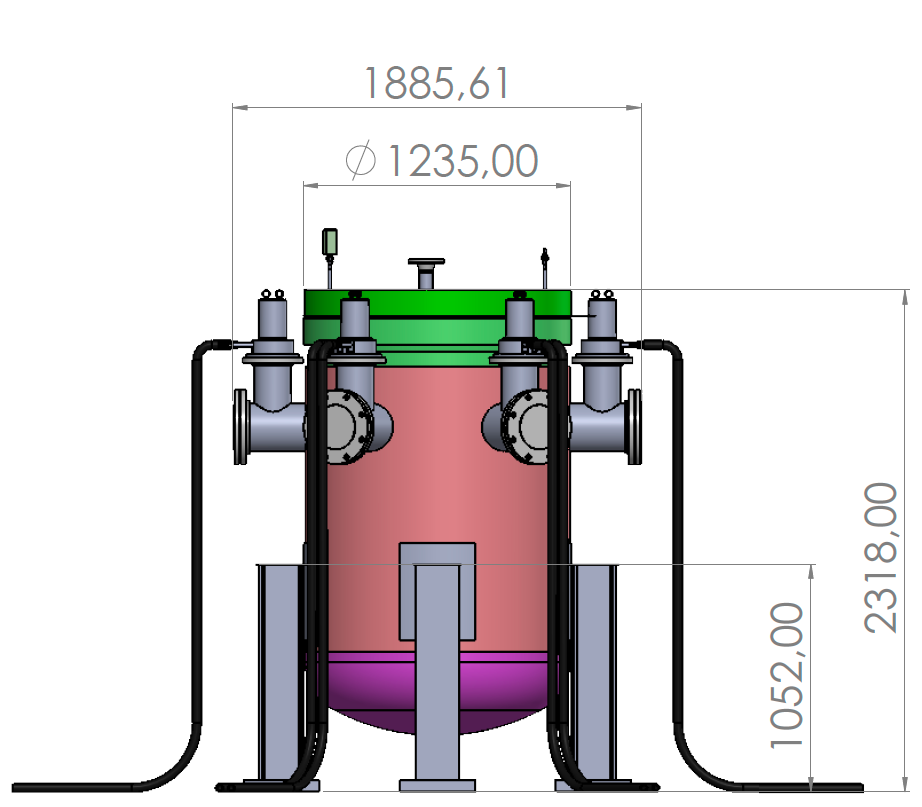}\hfill
    \includegraphics[width=0.48\textwidth]{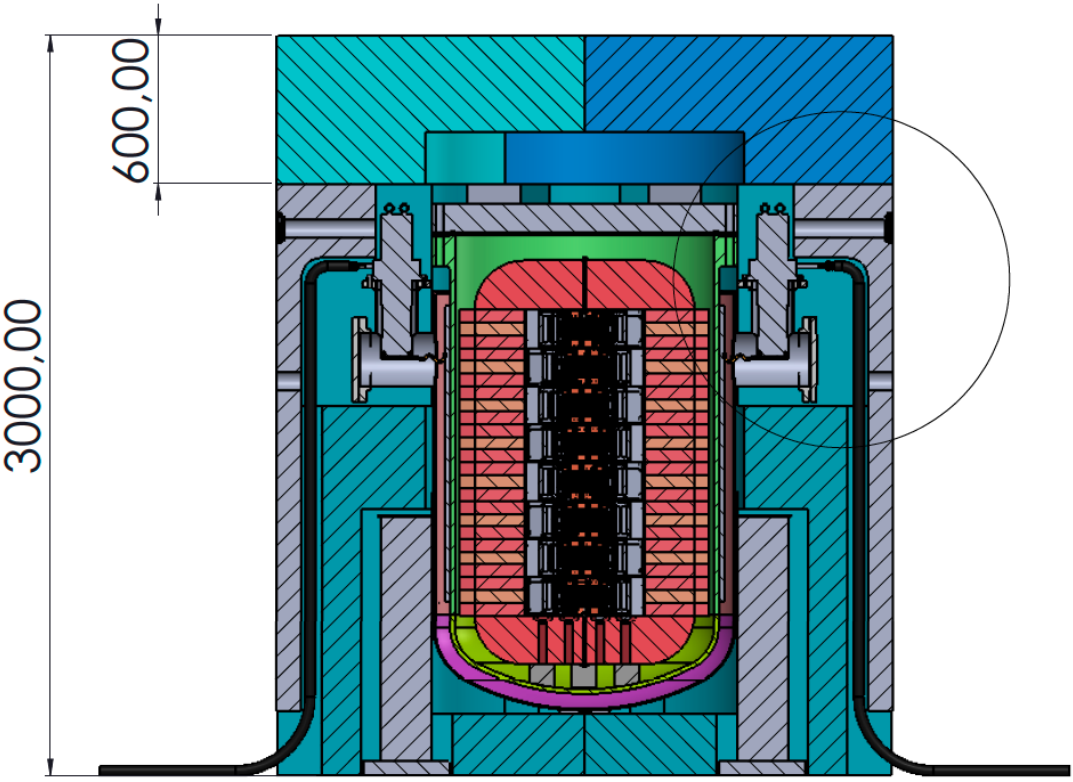}
    \caption{Left) Design of the Oscura pressure vessel for the operation of the 26~gigapixel skipper-CCD detector array. Right) Cross section of the Oscura vacuum vessel showing the internal lead and copper shield (dark/light pink), the external high-density polyethylene shield (dark/light blue), and the region filled with LN2 (green).
    }
    \label{fig:10kgsketch}
\end{figure}

“Dark current," i.e., thermal fluctuations of electrons from the valence to the conduction band, presents an irreducible source of 1$e^-$ events in skipper-CCDs. Operating the sensors with a low dark current requires cooling down the system to between 120K and 140K (the optimal operating point will be determined from the prototype sensors). The current strategy for the cooling system is to submerge the full detector array in a Liquid Nitrogen (LN2) bath operated with a vapor pressure of 450 psi to reach this temperature. Closed-cycle cryocoolers will provide the full system cooling capacity (less than 1~kW power)~\cite{AL600}. A schematic of the pressure vessel and its radiation shield is shown in Fig.~\ref{fig:10kgsketch}.

\section{Instrumental background sources in Oscura skipper-CCDs} \label{sec:instbkgs}
The two main contributions to the Oscura background rate come from radiation and instrumental background sources. Oscura plans to reach a radiation background of 0.01 dru using strict background control techniques and shielding. However, instrumental sources of events with few electrons (2$e^-$, 3$e^-$,..., 10$e^-$) must also be addressed. In this section, we discuss these background sources. Based on the previously defined Oscura background goal/requirement, we establish a set of constraints on the performance parameters of the Oscura sensors. This is summarized in Table~\ref{tab:reqs}.

\subsection{Thermal dark current} \label{sec:reqdc}
Thermal dark current is an irreducible source of 1$e^-$ events in skipper-CCDs and constrains the lowest 1$e^-$ rate ($R_{1e^-}$) that can be achieved by Oscura. The 1$e^-$ events coming from dark current will generate pixels with 2$e^-$ or more by accidental coincidences. The count of $ne^-$ single pixel events for the 30~kg-year Oscura exposure can be calculated assuming a Poisson distribution,
\begin{equation}
    K_n = \frac{\lambda^n e^{-\lambda}}{n!}  \times N_{pix}  \times N_{exp} \times (365 \times  3) , \label{eq:kn}
\end{equation}
where $N_{exp}$ is the total number of exposures per day, $N_{pix}$ is the total number of pixels in Oscura, and we assume a 3-year data-taking run. Here, $\lambda = R_{DC, 1e^-}/N_{exp}$, where $R_{DC, 1e^-}$ is the mean 1$e^-$ rate coming from dark current in units of $e^-$/pix/day. In Table~\ref{tab:accHits} we present $K_n$ for different running conditions defined by $R_{DC, 1e^-}$ and $N_{exp}$.
\begin{table}[h!]
\caption{Counts of 2$e^-$, 3$e^-$, and 4$e^-$ single pixel events generated by accidental coincidences from the thermal dark current. $N_{exp}=1$~(12)~exposure(s)/day means that the full readout of the detector takes 24~(2) hours and 26 gigapixels are assumed for the 10~kg array.}\label{tab:accHits}
	\centering	
	\small
        \vspace{0.25cm}
	\begin{tabular}{|l|l|l|l|}
	\hline
		Run conditions              & $2e^-$ & $3e^-$ & $4e^-$ \\ \hline
		$R_{DC, 1e^-}=1.6\times10^{-4}$ &        &        &        \\
      	\quad \quad $N_{exp}$=1	    & 364k   & 19     & 0      \\
      	\quad \quad $N_{exp}$=12    & 30k    & 0.1    & 0      \\
       \hline
 		$R_{DC, 1e^-}=1\times10^{-5}$   &        &        &        \\
      	\quad \quad $N_{exp}$=1	    & 1.4k   & 0      & 0      \\
      	\quad \quad $N_{exp}$=12    & 119    & 0      & 0      \\
       \hline
            $R_{DC, 1e^-}=1\times10^{-6}$   &        &        &        \\
      	\quad \quad $N_{exp}$=1	    & 14.2   & 0      & 0      \\
      	\quad \quad $N_{exp}$=12	& 1.2    & 0      & 0      \\
        \hline
	\end{tabular}
\end{table}

Table~\ref{tab:accHits} shows that, in order to avoid accidental coincidences, it is better to have more exposures per day (large $N_{exp}$). The lowest $R_{1e^-}$ achieved in skipper-CCD detectors, reported by SENSEI~\cite{SENSEI:2020dpa}, is $1.6 \times 10^{-4}$ $e^-$/pix/day. Assuming we achieve this rate and considering dark current as its origin, we will have less than one accidental 3$e^-$ events, achieving the Oscura background requirement, if we operate with 2-hour exposures, i.e. $N_{exp}=12$~exposures/day. The Oscura sensors would still comply with the requirement with SENSEI's $R_{1e^-}$ using 5-hour exposures. Table~\ref{tab:accHits} also indicates that we need to improve the SENSEI rate by at least two orders of magnitude and read out the full detector in less than 2 hours to have less than one accidental coincidences of 2$e^-$ in one pixel, the Oscura background goal.

The Oscura CCDs are sensors with 1.35 Mpix. This means that, for each CCD, we need a readout rate higher than 188 (76)~pix/s to reach Oscura background goal (requirement). As we plan to read each sensor with a single amplifier, the pixel readout time should be less than 5.3 (13.1)~ms.

\subsection{Readout noise} \label{sec:reqnoise}
The readout noise and the threshold used to determine if a pixel has $ne^-$ define the number of $(n-1)e^-$ single pixel events that fall above the threshold to be counted as $ne^-$ events. In principle, a skipper-CCD's readout noise can be made extremely small when multiple skipper samples ($N_{skp}$) are collected, as it drops as $1/\sqrt{N_{skp}}$~\cite{skipper2017}. However, adding skipper samples makes the readout slower and, as shown in Table~\ref{tab:accHits}, longer readout times produce more accidental $ne^-$ events. An optimization between the readout noise and speed should then be considered when choosing $N_{skp}$.

The total count of $(n-1)e^-$ single pixel events counted as $ne^-$ events comes from integrating the tail of the $(n-1)e^-$ single pixel event normal distribution from the threshold for counting $ne^-$ events. It is given by
\begin{equation} \label{eq:ln}
L_n = \frac{1}{2}\left[1-\erf\left(e_{th}/\sqrt{2}\sigma_{noise}\right)\right] K_{(n-1)}, 
\end{equation}
where $K_{(n-1)}$ is the total number of $(n-1)e^-$ single pixel events (see Eq.~\eqref{eq:kn}), $\erf$ is the error function, $\sigma_{noise}$ is the electronic readout noise in units of electrons, and $(n-1)+ e_{th}$ is the threshold used to determine if a pixel has $ne^-$. For example, for $n=2$, $e_{th}=0.5$, if the threshold is set to 1.5$e^-$.

From Eq.~\eqref{eq:ln} we see that if $R_{DC, 1e^-}=1 \times 10^{-6}\,(1.6\times10^{-4})$~$e^-$/pix/day, we need $e_{th}/\sigma_{noise}> 5.4\,(4)$ in order to get $L_{2\,(3)}<1$, consistent with the Oscura background goal (requirement). As the noise increases, we need to increase $e_{th}$ to keep $L_n<1$, but higher values of $e_{th}$ produce inefficiency for counting $ne^-$ single pixel events. In fact, the efficiency (eff) is the integral of the $ne^-$ single pixel event normal distribution from the given threshold, and can be calculated as
\begin{equation}
\textrm{eff} = \frac{1}{2}\left[1+\erf\left((1-e_{th})/\sqrt{2}\sigma_{noise}\right)\right]. \label{eq:eff}
\end{equation}

\begin{figure}[h!] 
    \centering
    \includegraphics[width=0.5\linewidth]{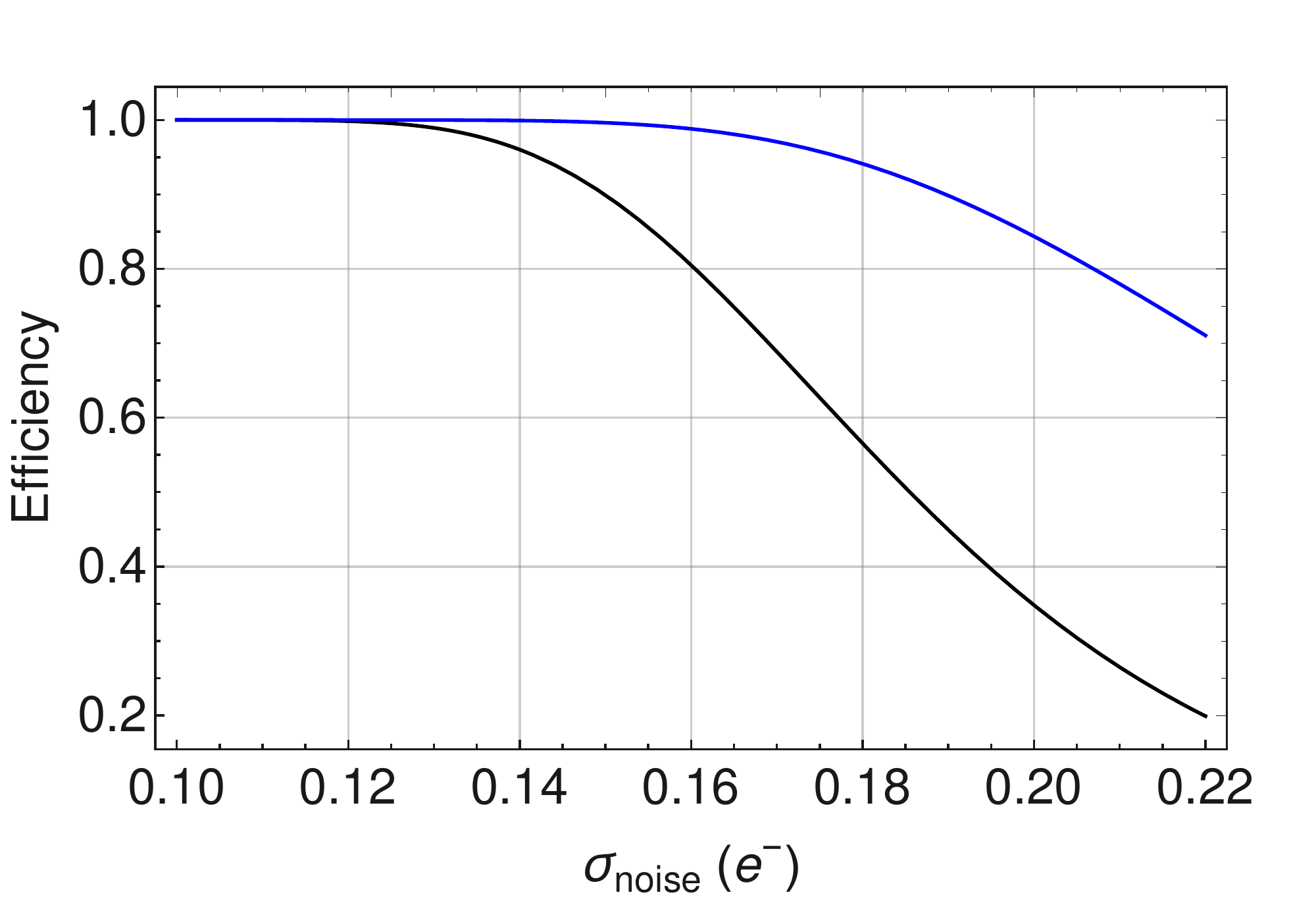} \hfill
    \caption{Efficiency for counting $2\,(3)\,e^-$ single pixel events as a function of the electronic readout noise in black (blue) when the threshold is set such that the Oscura background goal (requirement) is achieved for $R_{DC, 1e^-}=1\times 10^{-6}\,(1.6\times10^{-4})\,e^-$/pix/day.}
    \label{fig:ethnoise}
\end{figure}

From Eq.~\eqref{eq:eff} we see that $e_{th} = 1$ corresponds to 50\% efficiency, independent of the value of $\sigma_{noise}$. Fig.~\ref{fig:ethnoise} shows the efficiency for counting $2\,(3)\,e^-$ events as a function of the readout noise after imposing the conditions $e_{th}/\sigma_{noise}> 5.4\,(4)$. Based on this analysis, to maintain an efficiency for counting $2\,(3)\,e^-$ events higher than 80\% while complying with the Oscura background goal (requirement) we need $\sigma_{noise}<0.16\,(0.20)\,e^-$.

\subsection{Spurious charge} \label{sec:reqsp}
The high electric field generated in the CCD when the gate voltages change to move the charge from one pixel to the other can lead to the production of spurious charge (SC), also known as clock-induced charge~\cite{Janesickbook2001}. This can happen either in the active area or in the serial register and it strongly depends on the clock rise time and the clock swings. The primary source of spurious charge is the clocking of the serial register, which tends to dominate over the slower vertical clocks due to the higher capacitance of the line across the CCD active region. Assuming that the probability of generating one electron on a single pixel transfer, $\kappa_{SC}$, is the same in both registers and considering that each pixel on the CCD is shifted $N_{ser}$ ($N_{par}$) times in the serial (parallel) register during readout, the 1$e^-$ rate coming from the spurious charge is
\begin{equation} \label{eq:sp}
R_{SC, 1e^-} = N_{exp} \times (N_{ser}+N_{par}) \times \kappa_{SC}.
\end{equation}

Oscura requires this component of the 1$e^-$ rate to be subdominant to the thermal dark current, $R_{SC, 1e^-} < R_{DC, 1e^-}$. Assuming $R_{DC, 1e^-}=1\times10^{-6}\,(1.6\times10^{-4})\,e^-$/pix/day, $N_{exp}=12$~exposures/day, $N_{ser}=1058$~transfers/exposure, and $N_{par}=1278$~transfers/exposure, results in the condition $\kappa_{SC}<4\times10^{-11}\,(6\times10^{-9})\,e^-$/pix/transfer to reach the Oscura background goal (requirement).

\subsection{Traps} \label{sec:reqtraps}
Defects within the silicon lattice create intermediate energy levels within the Si bandgap that act like traps. These traps usually capture one electron from charge packets as they are transferred through the device and release the charge at a later characteristic time $\tau$ dependent on the temperature.

The number of 1$e^-$ events per sensor per exposure coming from traps is $N_{hits} \times N_{traps}$, where $N_{hits}$ is the number of hits, i.e., pixels with more than $2e^-$, in one exposure and $N_{traps}$ is the mean number of traps that a hit traverses during readout, which equals the total number of electrons trapped per hit assuming that each trap captures one electron. Only traps with a $\tau$ larger than the pixel readout time are considered because faster traps will release the trapped electron in the pixel containing the hit. The rate of $1e^-$ events produced from traps is calculated as
\begin{equation}
R_{T, 1e^-} =  \frac{N_{hits} \times N_{traps}}{N_{pix}} N_{exp}\,,
\end{equation}
where $N_{pix}$ is the total number of pixels in each sensor.

We estimate $N_{hits}$ for Oscura assuming the baseline 1.35 Mpix sensors with 0.5 g of active mass and a background rate of 0.01 dru. In these conditions, we expect $N_{events}=5 \times 10^{-4}$~events/exposure/sensor up to 100 keV in a one-day exposure, i.e., $N_{exp}= 1$~exposure/day. Then, $N_{hits}=n_{pix}\times N_{events}$, where $n_{pix}$ is the expected number of pixels in one event with energy below 100~keV. As the number of pixels in one event is a broad distribution that increases towards less number of pixels, a conservative assumption is to take $n_{pix}=10$~pix/event. This results in
\begin{equation} \label{eq:traps}
R_{T, 1e^-} =  (3.7 \times 10^{-9}~\mbox{hits/pix/day}) \times N_{traps}.
\end{equation}

We require $R_{T, 1e^-}<R_{DC, 1e^-}=1\times10^{-6} e^-$/pix/day to achieve both the Oscura background goal and requirement. This imposes the condition $N_{traps} < 2.7\times10^2$. Considering that each hit traverses a maximum of $N_{ser}+N_{par}=2336$ pix, the allowed density of traps is
\begin{equation}
\rho_{traps} = \frac{N_{traps}}{N_{ser}+N_{par}} < \frac{2.7\times10^2}{2336}\simeq0.12 \mbox{ traps/pix}.
\end{equation}
This condition is satisfied if there is a trap every $\sim$8~pixels. Note that the allowed density of traps depends inversely on the background rate. Assuming a background rate one order of magnitude higher than the expected for Oscura, i.e. 0.1 dru, we get $\rho_{traps}<0.012$~traps/pix.

\subsection{Charge transfer inefficiency}
Charge transfer inefficiency (CTI) refers to the loss of charge when a charge packet is moved from one pixel to the next. It depends on several different parameters such as trap populations, trap densities, clocking time, clocking sequence, and temperature~\cite{Janesickbook2001, Townsley2002, Grant2006, Kanemaru2020}. CTI in the Oscura CCDs will result in the misidentification of $ne^-$ single pixel events into $(n-1)e^-$ events. The fraction of a $ne^-$ event in a single pixel that will be left in the subsequent pixel due to CTI is
\begin{equation} \label{eq:cti}
\varepsilon_{CTI} = k_{CTI} (N_{ser}+N_{par}),
\end{equation}
where $k_{CTI}$ is the charge transfer inefficiency measured for a single pixel transfer within a row/column. Here, we are assuming a similar inefficiency for serial and parallel transfers. CTI is not an impediment to reach Oscura background goal/requirement as it does not change the total event rate. However, for a good performance, consistent with what is commonly achieved in CCDs, we establish as a target $\varepsilon_{CTI}< 0.01$, which means $k_{CTI}<5\times10^{-6}$.

\subsection{Light generation in LN2 and other detector materials}\label{sec:reqlight}
As discussed above, the skipper-CCDs for Oscura will be operated in a LN2 pressure vessel. For a run on the surface and without any shield, we measure light generated in LN2 at a rate of $R_{LN2, 1e^-}=0.013~e^-$/pix/day using a SENSEI skipper-CCD. The light is assumed to be produced by environmental radiation interacting in the LN2. For this measurement, the background around 10 keV was $\sim$$10^4$~dru, six orders of magnitude above the Oscura radiation background target of 0.01 dru. Assuming that light generation in LN2 scales with the background rate at higher energies, we estimate this light to produce $R_{LN2, 1e^-}^{0.01\textrm{dru}}\sim$$10^{-8}~e^-$/pix/day in Oscura. This is much less than the expected thermal dark current in the Oscura sensors and it is not expected to contribute to the experimental background. We will check this simple assumption in next iterations of our experiment.

However, since the geometry and CCD packaging used to measure light generation in LN2 are not identical to the planned Oscura design, we are working to implement a light shield to ensure that ionization events from visible and near-IR light are a subdominant background. We aim to suppress more than $90$\% of the light hitting the surface of the Oscura sensors.

\section{Oscura prototype sensors performance} \label{sec:perf}
Before Oscura, skipper-CCDs for DM experiments were fabricated at a 150~mm diameter wafer foundry that is in the process of discontinuing the CCD processing line. The development of large-scale CCD fabrication techniques in partnership with new foundries was identified as the main Oscura technical risk. We have successfully overcome it, developing a fabrication process for Oscura skipper-CCDs on 200~mm diameter wafers with a new industrial partner (Microchip Technology Inc.) and also with a government laboratory (MIT-LL).

\begin{figure}[h!]
    \includegraphics[width=0.47\linewidth]{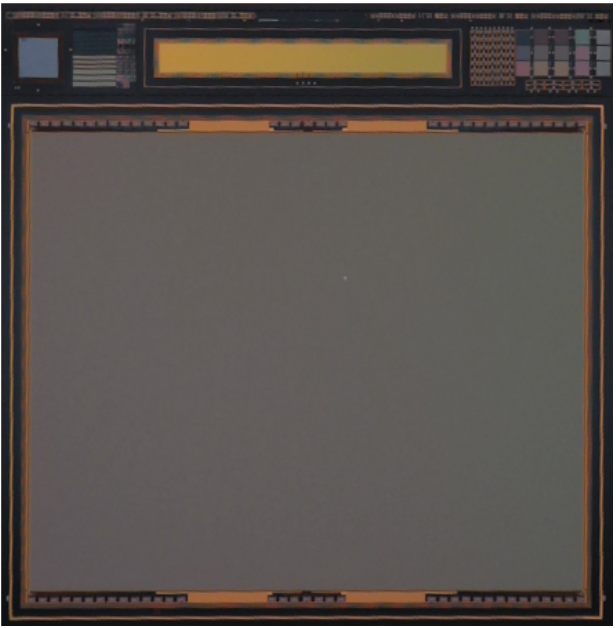} \hfill
    \includegraphics[width=0.47\linewidth]{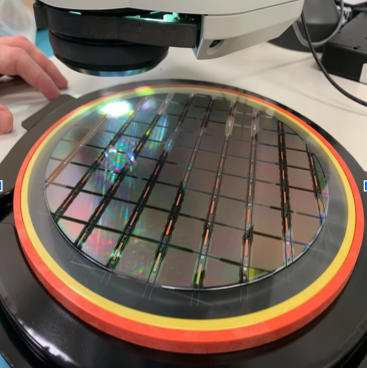}
    \caption{Left) Skipper-CCD fabricated for Oscura at Microchip in 2021 (design from S. Holland - LBNL). The upper structures in the picture are for performing tests. Right) 200~mm diameter wafer with $\sim$50 skipper-CCDs fabricated for Oscura at Microchip.}\label{fig:microchip}
\end{figure}

The overall design of the Oscura prototype sensors is very similar to that of the skipper-CCDs used in the SENSEI~\cite{sensei2019} and DAMIC-M~\cite{2020DAMICM} experiments. Oscura skipper-CCDs are small format sensors, with $1278\times1058$ pixels, and 4 skipper-CCD amplifiers, one in each corner. The new three-phase skipper-CCDs have been fabricated in 200~mm diameter wafers, using high-resistivity silicon wafers as a starting material. Previous skipper-CCD experiments have used a similar starting material. Fig.~\ref{fig:microchip} shows pictures of an Oscura prototype skipper-CCD (left) and a 200~mm diameter wafer with $\sim$50 Oscura sensors (right).

In this section, we present the performance of the first fabricated skipper-CCDs for Oscura. We compare it to the constraints discussed in Section~\ref{sec:instbkgs} and we discuss the strategy to control the instrumental background sources. This is summarized in Table~\ref{tab:reqs}. Most of the tests presented here were done using individual Oscura prototype skipper-CCDs packaged in copper trays and installed in dedicated testing setups at the Silicon Detector Facility, at the Fermi National Accelerator Laboratory (FNAL).

\subsection{Readout noise and speed} \label{sec:perfnoise}
\begin{figure}[h!]
\centering
    \includegraphics[width=0.47\textwidth]{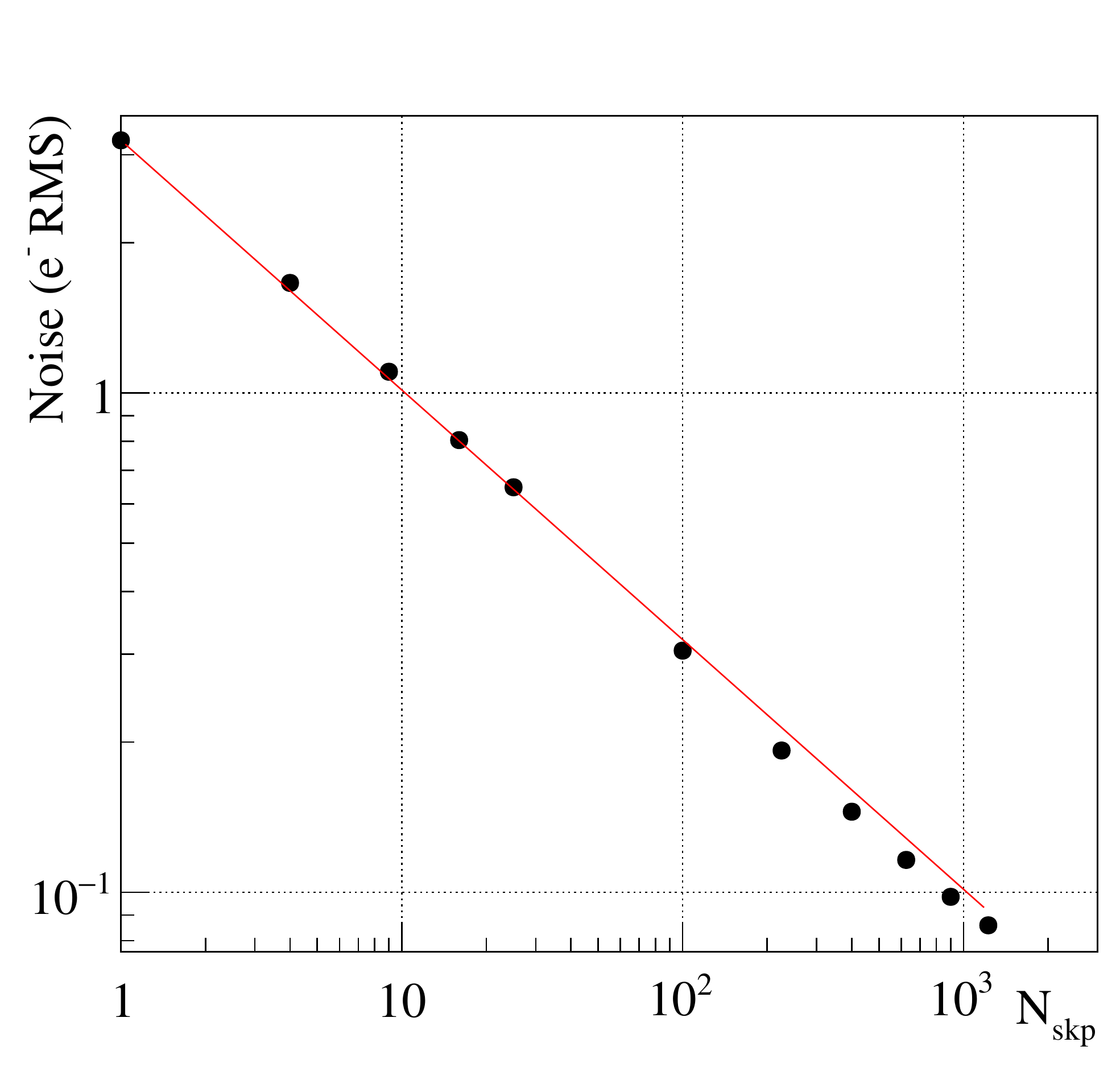} \hfill
    \includegraphics[width=0.47\textwidth]{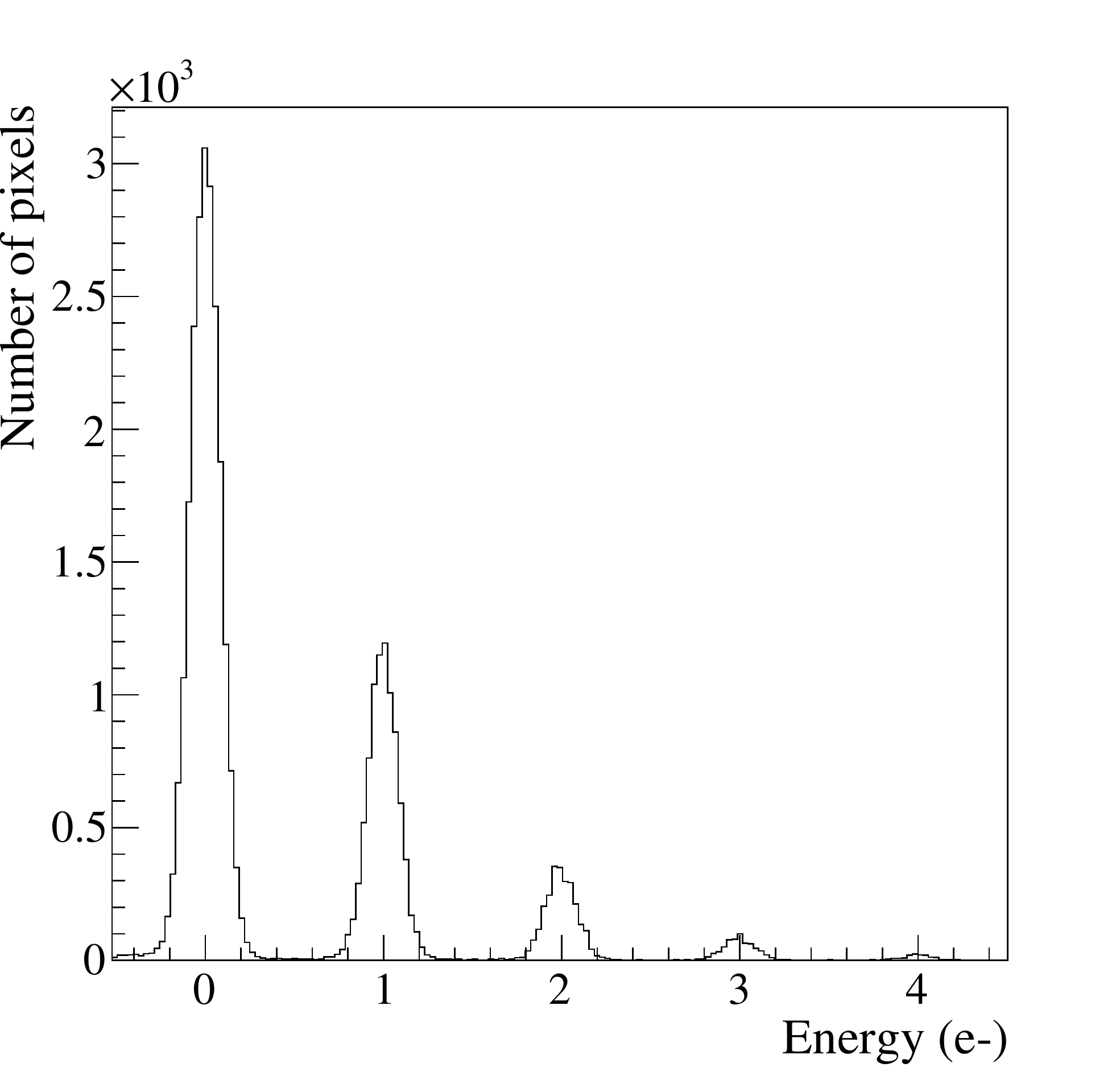}
    \caption{Left) Individually packaged Oscura prototype skipper-CCD readout noise as a function of $N_{skp}$. The expected $1/\sqrt{N_{skp}}$ dependence is shown in red. Right) Charge pixel distribution from acquisition with $N_{skp}=1225$ samples per pixel; the electron-counting capability of the first fabricated Oscura prototype skipper-CCDs was demonstrated with this result, see~\cite{Cervantes2023}.}\label{fig:noise}
\end{figure}

Using an individually packaged Oscura prototype sensor, we measure the readout noise as a function of $N_{skp}$. The results, shown in Fig.~\ref{fig:noise}, demonstrate that $N_{skp}=400\,(225)$ are enough to reach a noise of $0.15\,(0.19)\,e^-$, consistent with the constraints to achieve Oscura background goal (requirement) discussed in Section~\ref{sec:reqnoise}. However, increasing $N_{skp}$ also increases the readout time and the constraint on this parameter should also be met to comply with the Oscura background goal/requirement. In these measurements, the pixel readout time for $N_{skp}=400\,(225)$ was 15.3 (9)~ms. This corresponds to a pixel readout rate of 65 (111)~pix/s, allowing to read out the whole array in 5.8 (3.4)~hours. Then, the readout time and noise constraints are both met only for the background requirement.

\begin{figure}[h!]
	\includegraphics[width=1.\linewidth]{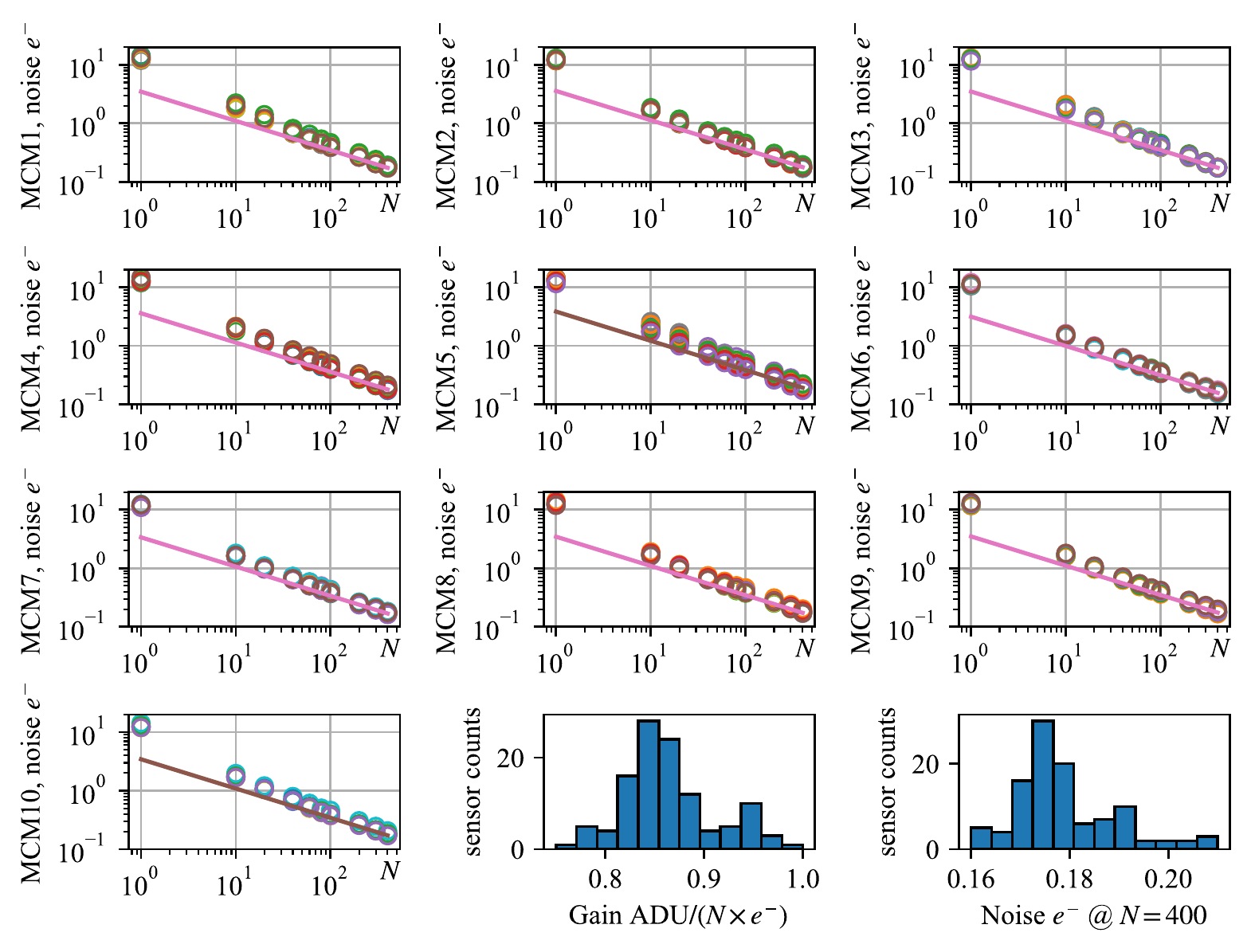}
	\caption{Noise as a function of $N_{skp}$ for all 10 MCMs and channels. The expected $1/\sqrt{N_{skp}}$ dependence is shown in pink, consistent with the noise performance for $N_{skp}>20$. For $N_{skp}\leq20$, the noise is not dominated by the CCD performance but by the analog readout electronics~\cite{Chierchie2023, Sofo2021}. The blue histograms correspond to the gain and noise distribution computed for $N_{skp}=400$. Taken from~\cite{Chierchie2023}.}\label{fig:noiseMCM}
\end{figure}

Tests were also performed in a system designed to host 16 MCMs, but with 10 MCMs installed. As mentioned in Section~\ref{sec:design}, each MCM has 16 Oscura prototype skipper-CCDs. Details and results from measurements with this system can be found in Ref.~\cite{Chierchie2023}. Fig.~\ref{fig:noiseMCM} shows the readout noise as a function of $N_{skp}$ for all the skipper-CCDs in the 10 MCMs. From these results, with $N_{skp}=480\,(300)$ the system reaches a noise of $0.16\,(0.20)\,e^-$~RMS, consistent with the constraints to achieve Oscura background goal (requirement) discussed in Section~\ref{sec:reqnoise}. In this system, the pixel readout time for $N_{skp}=480\,(300)$ is $t_{pix}=16.8\,(10.5)$~ms, plus an additional $t_{mux}=0.64$~ms for multiplexing the 16 MCMs. This corresponds to a pixel readout rate of 57 (89)~pix/s, allowing to read out the whole array in 6.6 (4.2)~hours. Again, the readout time and noise constraints are both met only for the Oscura background requirement. The system is yet to be optimized in its final configuration, which will enable it to achieve a higher pixel readout rate. The current system is still missing the MIDNA ASIC (Application Specific Integrated Circuit)~\cite{MIDNA}, which will perform the analog pixel processing, and the flex cables used in this setup are longer than in the Oscura design. With shorter cables and the ASIC, the system will produce a faster signal due to reduced capacitance and the higher bandwidth of the MIDNA ASIC. This allows a reduction of the dead times in the readout sequence to get the maximum noise reduction per unit of readout time. The multiplexing time is also expected to be reduced using a faster ADC (Analog to Digital Converter) stage.

Note that we could achieve the readout rate necessary to reach the Oscura background goal by performing on-chip binning, i.e. combine the charge of adjacent pixels, during the readout, at the cost of reducing spatial resolution.

\subsection{Dark current and single electron rate}~\label{sec:perfdc}
\begin{figure}[h!]
\centering
    \includegraphics[width=0.47\linewidth]{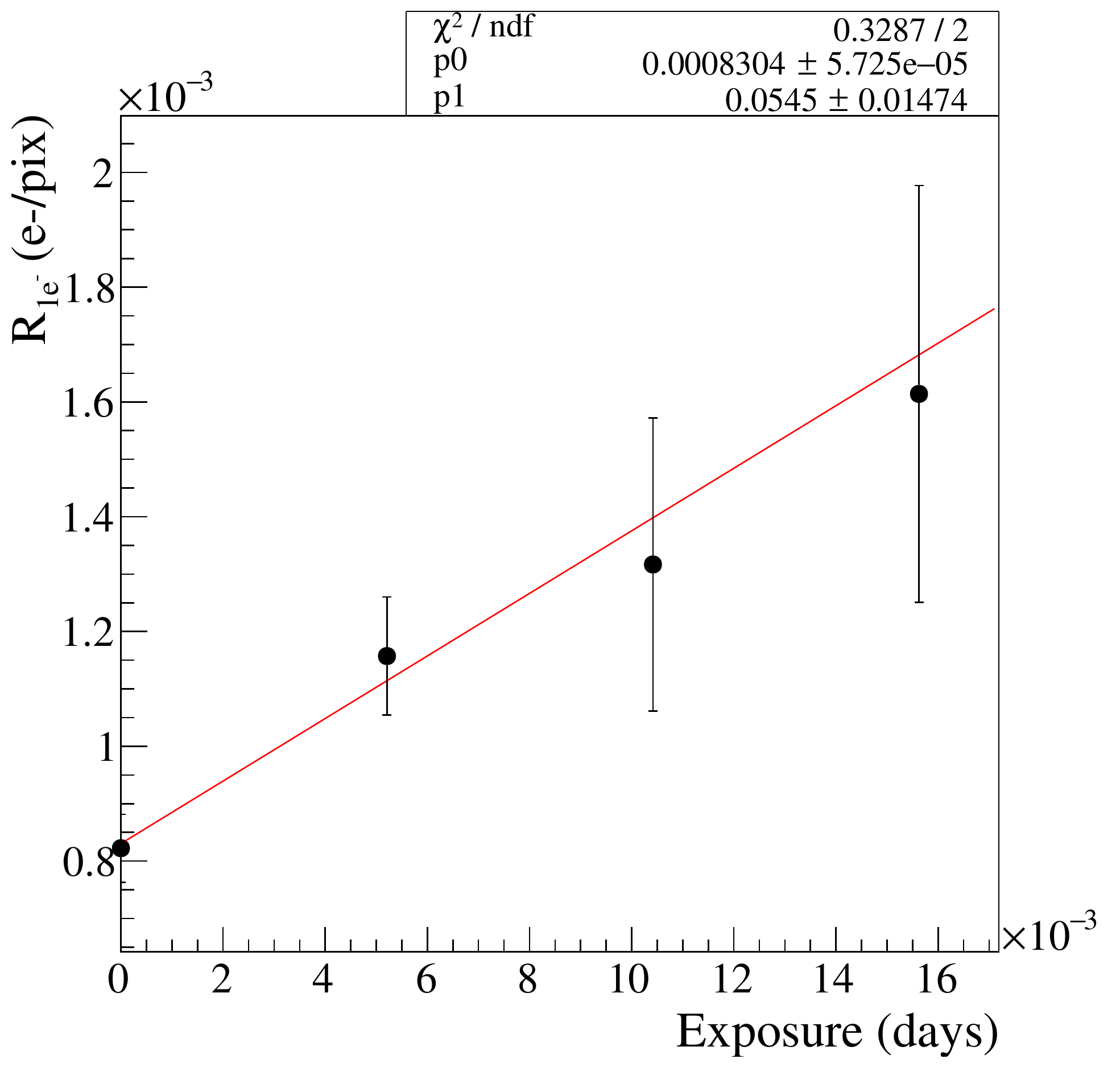}\hfill
    \includegraphics[width=0.47\linewidth]{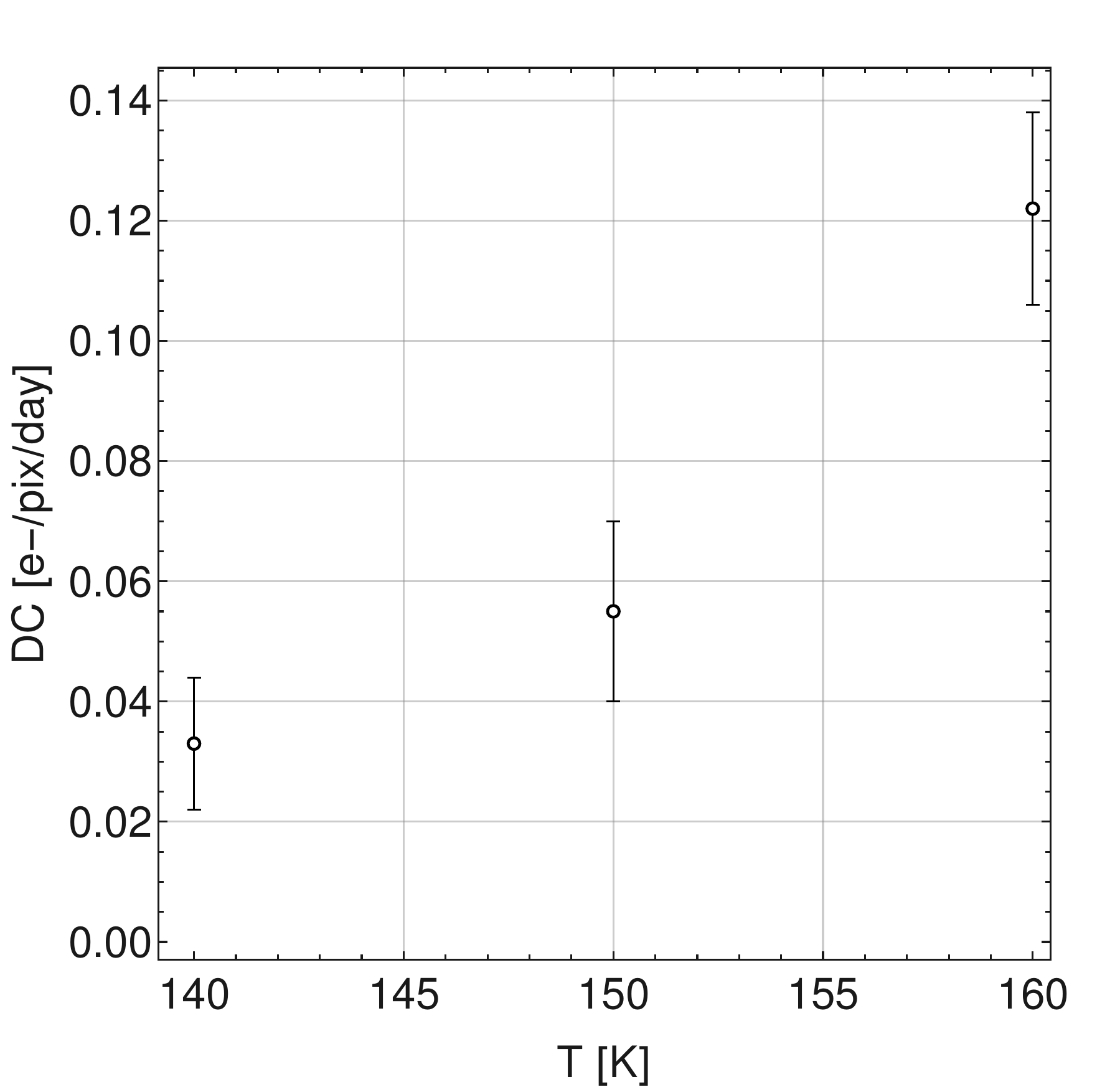}
    \caption{Left) $1e^-$ event rate as a function of exposure time for images taken at T=150K. The linear fit is shown in red. Right) Dark current (DC) measurements as a function of temperature at the surface with an individually packaged Oscura prototype skipper-CCD.}\label{fig:darkcurrent}
\end{figure}

To quantify the Oscura sensors dark current (DC), we measured the exposure-dependent $1e^-$ rate as a function of temperature with an individually packaged Oscura prototype skipper-CCD in a dedicated setup with 2~inches of lead shield at surface. At a given temperature, we acquired images with different exposure times, from 0 to 30~min, with $N_{skp}=200$. To increase the readout rate, we performed a $5\times1$ binning, i.e., the charge of 5 consecutive pixels in the same row was summed before readout. For the analysis, we selected the first rows of each image that were free of high-energy events. We perform linear fits to the plots of $1e^-$ rate as a function of exposure time, where the slopes correspond to the exposure-dependent $1e^-$ event rate. Fig.~\ref{fig:darkcurrent} (left) shows one of these plots corresponding to images taken at $T=150~\mathrm{K}$. We performed this measurement at different temperatures and the results, first presented in~\cite{Cervantes2023}, are shown in Fig.~\ref{fig:darkcurrent} (right). The lowest value achieved was 0.03~$e^-$/pix/day, at 140~K.

This result is much larger than the rate needed to achieve the Oscura background requirement/goal. However, it is well known that, at surface, the main contribution to the $1e^-$ rate for $T<150~\mathrm{K}$ does not come from dark current but from low-energy radiation that is created when high-energy events interact with the detector components~\cite{SENSEI:2020dpa, Du:2020ldo, Moroni2022}. At surface, where the rate of events with energies above 0.5~keV is $\mathcal{O}\left(10^4-10^5\right)$, we expect a $\mathcal{O}\left(10^{-2}\right)$ $1e^-$ rate, consistent with the computation done in Ref.~\cite{Du:2020ldo} for the SENSEI at MINOS setup, where a $\mathcal{O}\left(10^{-4}\right)$ $1e^-$ rate is expected from the $\sim$3~kdru high-energy background rate. Lower exposure-dependent $1e^-$ rates are expected when measurements can be made underground, in lower-background environments.

\subsection{Spurious charge}~\label{sec:perfsp}
From the measurements described in Section~\ref{sec:perfdc}, we extract an upper limit for the generated spurious charge from the y-intercept of the linear fits. The average value is $8.4\times10^{-4} e^-$/pix. Considering that in these measurements each read pixel underwent $(N_{par}+N_{ser})/2=1168$~transfers/exposure, this corresponds to $\kappa_{SC}=7.2\times10^{-7} e^-$/pix/transfer.

Following an analogous procedure as the one described in~\cite{sensei2022}, we measured the generation of charge in the output stage. Using an individually packaged Oscura prototype sensor, we read out 20 pixels with 5 million skipper samples, clocking only the output stage. We obtained an average rate of $\sim$$1\times10^{-7} e^-$/sample.

With the SENSEI skipper-CCD used to measure light generation in LN2 we also computed the generated spurious charge. The data shows that the total number of $1e^-$ events produced by the spurious charge per pixel is $\sim$$10^{-4}$. As in this CCD $(N_{par}+N_{ser})\simeq7000$, this gives $\kappa_{SC} \simeq 1.4\times10^{-8} e^-$/pix/transfer.

In all cases we measure a $\kappa_{SC}$ higher than what is needed to meet Oscura background requirement/goal. Members of the Oscura collaboration are working to better understand spurious charge generation in skipper-CCDs. To reduce it, we are considering several approaches, including the use of filtering techniques to decrease the slew rates of horizontal clock signals, as well as implementing shaped clock signals~\cite{2018RogerSmith}.

\subsection{Trap density}
We use the charge pumping technique~\cite{Blouke1988, Janesickbook2001, Mostek2010, Hall2014} to localize and characterize traps in individually packaged skipper-CCDs. This popular method consists of filling the traps and allowing them to emit the trapped charge in their neighbor pixel multiple times. This is done by repeatedly moving, back and forth between the phases\footnote{A pixel phase refers to a gate (electrode) laying on the CCD front surface in which clocking voltages are applied~\cite{Janesickbook2001}.} in one pixel, a uniform illuminated field creating "dipole" signals relative to the flat background.

Using a violet LED externally controlled by an Arduino Nano, we uniformly illuminated the three-phase skipper-CCDs and performed a charge pumping sequence that probes traps below pixel phases 1 and 3. We collected images varying the time that charge stayed below the pixel phases ($dt_{ph}$). We identified and tracked the position of each dipole in the set of images, computed its intensity as a function of $dt_{ph}$ and fitted it. From the fits, we extracted the characteristic release time $\tau$ for each of the found traps. We did this at different temperatures.

\begin{figure}[h!]
    \centering
    \includegraphics[width=1.\linewidth]{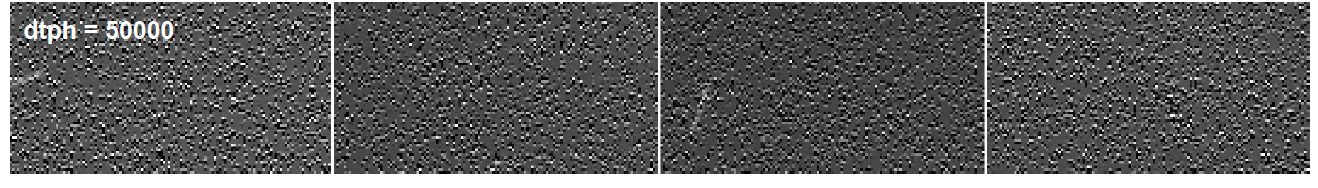}
    \includegraphics[width=1.\linewidth]{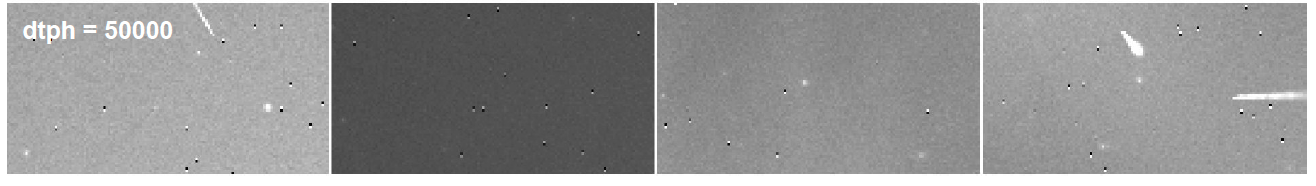}
    \caption{Section of images corresponding to each of the 4 amplifiers after performing pocket pumping with $dt_{ph}=50000$~clocks (3.34~ms) in Oscura prototype-A (top) and prototype-B (bottom), at 170K. The dipoles seen in the images correspond to charge traps under pixel phases 1 and 3.}\label{fig:qpump}
\end{figure}

Fig.~\ref{fig:qpump} shows images revealing traps with characteristic time $\tau>3.34$~ms, corresponding to $dt_{ph}=50000$~clocks, for two Oscura prototype sensors with different fabrication process. The four images in the top row, corresponding to each of the four amplifiers in prototype-A, show a much more significant density of dipoles compared to the images in the bottom row, corresponding to each of the amplifiers in prototype-B.

\begin{figure}[h!]
    \centering
    \includegraphics[width=0.75\linewidth]{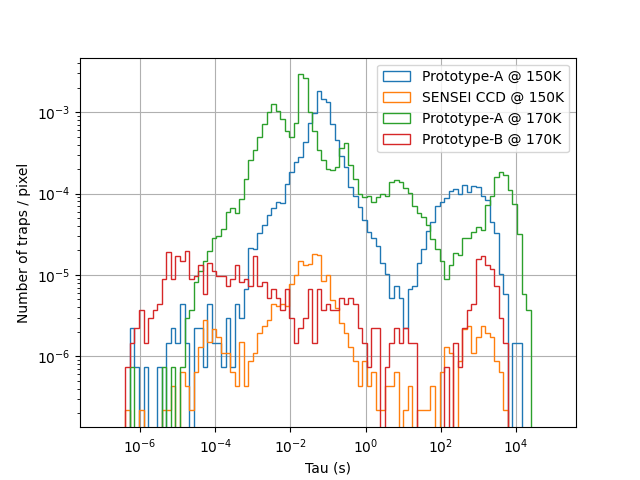}
    \caption{Number of traps per pixel as a function of the release characteristic time $\tau$ for: Oscura prototype-A at 150~K (blue) and at 170~K (green); Oscura prototype-B at 170~K (red); and a SENSEI skipper-CCD at 150~K (orange).}\label{fig:traps}
\end{figure}

The histograms in Fig.~\ref{fig:traps} show the number of traps per pixel as a function of $\tau$ for two different Oscura prototype skipper-CCDs and a SENSEI skipper-CCD, at two different temperatures. These histograms correspond to traps below pixel phases 1 and 3 and no detection efficiency was taken into account. Considering a uniform density of traps below the three phases in each pixel and assuming a conservative 10\% detection efficiency with a flat profile, the y-axis in Fig.~\ref{fig:traps} should be multiplied by a factor of 15 to obtain a more realistic trap density.

Traps with a release time greater than the pixel readout time will generate $1e^-$ events in the images. Considering the pixel readout rate needed to achieve Oscura background goal, traps with a release time $\tau>5.3$~ms should satisfy the trap density constraint. In the case of the Oscura prototype-A, the realistic number of traps with $\tau>5.3$~ms below 170~K is $\sim$$1.5\times10^{-2}$. For the Oscura prototype-B and the SENSEI skipper-CCD, this number is 2 orders of magnitude lower ($\sim$$3\times10^{-4}$). Despite this difference, both Oscura prototype skipper-CCDs meet the constraint needed to reach the background goal/requirement.

However, as discussed in Section~\ref{sec:reqtraps}, if Oscura overall background is one order of magnitude higher (0.1 dru), the Oscura prototype-A would barely meet the constraint. For this reason in a cooperative effort with the foundry that fabricated this sensor, we are trying to implement a different gettering method during the fabrication process to reduce possible impurities in the silicon.

\subsection{Charge transfer inefficiency}
\begin{figure}[h!]
    \centering
    \includegraphics[width=0.47\linewidth]{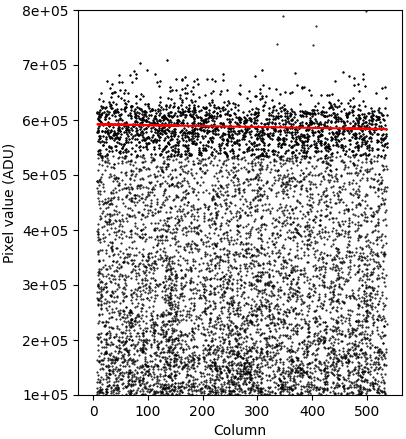}\hfill
    \includegraphics[width=0.47\linewidth]{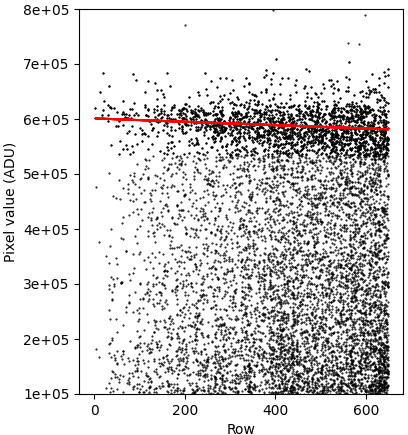}
    \caption{Scatter plots of the pixel values in analog-to-digital units (ADU) versus its column (left) and row (right) numbers for CTI measurements from an individually packaged Oscura prototype skipper-CCD exposed to a Fe$^{55}$ X-ray source. The linear fits to the X-ray pixel population are shown in red.} \label{fig:cti}
\end{figure}

We exposed an individually packaged Oscura prototype skipper-CCD to a Fe$^{55}$ X-ray source. We took images with 4 skipper samples applying the usual clocking sequence and voltages. We computed the parallel and serial registers CTI at different operational temperatures by linearly fitting the pixel population associated with X-ray depositions from X-ray transfer plots~\cite{Janesickbook2001}, see Fig.~\ref{fig:cti}. This figure shows the scatter plots of the pixel values versus its column (left) and row (right) numbers from a set of images acquired at 170~K. In all cases, we computed a $k_{CTI}<5\times10^{-5}$, which is slightly higher than the target. Note that these measurements were done using a Oscura prototype-A sensor with a high-density of traps and, as traps contribute to CTI, this number should be reduced by addressing this issue and optimizing the clocking sequence.

\subsection{Aluminum light shielding}
\begin{figure}[h!] 
	\includegraphics[width=1.0\linewidth]{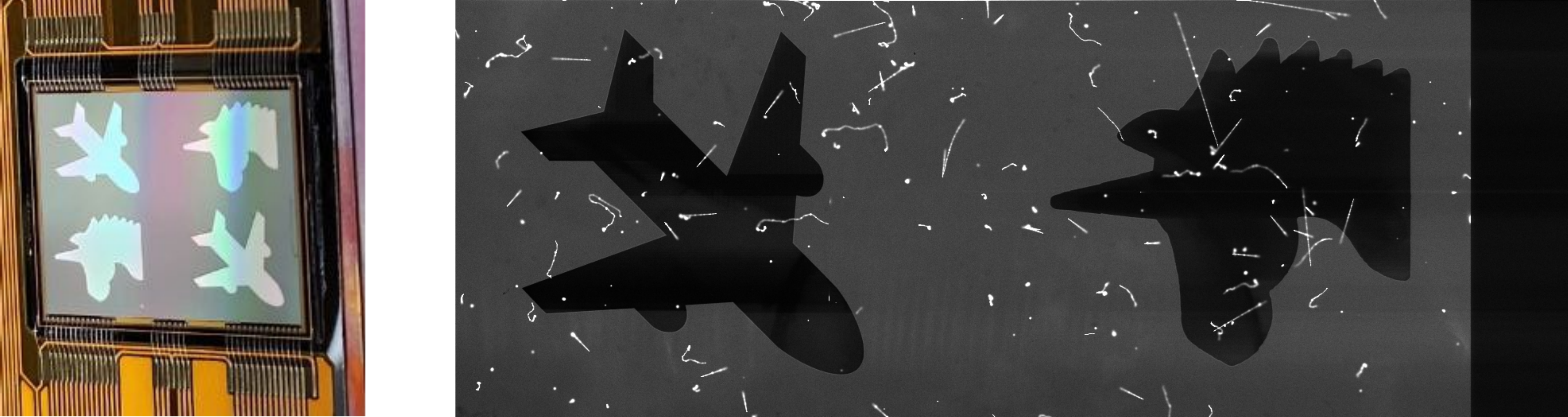}
	\caption{Oscura prototype skipper-CCD with an aluminum shield to probe the background light suppression potential of a 50\,nm metal layer. Left) Picture of the sensor with an aluminum plane- and unicorn-shaped layer on top of each quadrant. Right) Image acquired using the upper half of the Oscura skipper-CCD after 30 minutes exposure.}\label{fig:unicorn}
\end{figure}

We deposited a 50\,nm aluminum layer on top of the active area of Oscura prototype skipper-CCDs using a maskless lithography tool (Heidelberg MLA 150) and an electron beam evaporator (Temescal FC200). Within the first tests we produced a prototype with a shaped aluminum layer on top of each quadrant, as shown in Fig.~\ref{fig:unicorn} (left), where the wire bonds between the pads and the flex cable that connects the sensor to the readout electronics are also shown. Fig.~\ref{fig:unicorn} (right) shows an image taken with the upper half of the CCD after 30~min of exposure in a testing setup that was not completely shielded from environmental light. Electron, X-ray, and muon tracks are uniformly distributed in the active area, while the background light under the aluminum layer is $\sim$95\% suppressed compared to that in an uncovered area. Although the process implemented to produce this device is not optimal since the beam evaporator can damage the CCD, the result sufficed as proof of concept for the next fabrication step. A thicker aluminum layer can be safely incorporated as a part of the sensor production and will guarantee meeting the light suppression target.

\section{Summary and discussion} \label{sec:sumdisc}
In Table~\ref{tab:reqs}, we present the main instrumental sources of events with few electrons (2$e^-$, 3$e^-$,..., 10$e^-$) in Oscura skipper-CCDs, summarize the sensors performance parameters constraints to meet the Oscura background goal/requirement, discussed in Section~\ref{sec:instbkgs}, and show the quantified performance of the first fabricated Oscura prototype skipper-CCDs, presented in Section~\ref{sec:perf}, and the best performance achieved with skipper-CCDs. The trap density constraint is consistent with both the Oscura background goal and requirement. CTI and VIS/NIR light blocking targets are also presented in Table~\ref{tab:reqs}. For the full array readout time, the pixel readout rate and the readout noise, we show measurements from an individually packaged Oscura prototype skipper-CCD and, in brackets, from the system designed to host 16~MCMs. 
\begin{table}[h!]
\centering
\caption{Sensors performance parameters constraints to achieve the Oscura background goal/requirement and demonstrated performance of prototype sensors. The "Best achieved" column contains the best values achieved with skipper-CCDs and the checkmarks ($\checkmark$) indicate that the constraints for meeting the Oscura background requirement have been met.
}\label{tab:reqs}
\small
\vspace{0.25 cm}
\begin{tabular}{|l|l|l|l|l|l|l|}
\hline
Parameter              & Goal       & Requirement & Prototype & Best achieved & Units \\
\hline
Dark current           & $1\times10^{-6}$ & $1.6\times10^{-4}$\hspace{10pt} & $3\times10^{-2}$ & $1.6\times10^{-4}$\hfill$\checkmark$ & $e^-$/pix/day \\
Readout time (full array)  & $<2$ & $<5$ & 3.4 (4.2) & 3.4\hfill$\checkmark$ & hours \\
Pixel readout rate & $>188$ & $>76$ & 111 (89) & 111\hfill$\checkmark$ & pix/s  \\
Readout noise & $<0.16$ & $<0.20$ & 0.19 (0.20) & 0.19\hfill$\checkmark$ & $e^-$ RMS  \\
Spurious charge & $<4\times10^{-11}$ & $<6\times10^{-9}$ & $7.2\times10^{-7}$ & $1.4\times10^{-8}$ & $e^-$/pix/transfer \\
\cline{2-3}
Trap density ($\tau>5.3$~ms) & \multicolumn{2}{c|}{$<0.12$} & $<0.015$ & $<0.0003$\hfill$\checkmark$ & traps/pix \\
Charge transfer inefficiency  & \multicolumn{2}{c|}{$<10^{-5}$} & $<5\times10^{-5}$ & $<10^{-5}$\hfill$\checkmark$ & 1/transfer \\
VIS/NIR light blocking        & \multicolumn{2}{c|}{$>90\%$} & 95\% & 95\%\hfill$\checkmark$ & \\
\hline
\end{tabular}
\end{table}

From Table~\ref{tab:reqs}, we see that there are two main items where the Oscura prototypes fail to meet the constraints to achieve the background requirement: dark current and spurious charge; also, the CTI target has not been reached. Discussion on ways to improve the prototypes' performance in each of these items can be found in Section~\ref{sec:perf}. As a summary, we know that in the DC prototype measurement, dark current is not our main contribution; therefore, it should be taken as an upper limit as we expect our ultimate DC to be below $1.6\times10^{-4}$, the lowest $R_{1e^-}$ achieved in skipper-CCD detectors~\cite{SENSEI:2020dpa}. Also, the CTI target has already been reached with skipper-CCDs and it is expected to be achieved in Oscura prototypes with a low density of traps. Finally, the spurious charge constraint to meet the requirement has never been achieved with skipper-CCDs. Therefore, it is nowadays our biggest source of instrumental background.

As $\smash{R_{SC, 1e^-}}$ is proportional to the number of exposures taken per day, we can reduce its contribution by taking longer exposures (small $N_{exp}$). However, as discussed in Section~\ref{sec:reqdc}, with longer exposures the number of accidental coincidences from thermal dark current increases. Then, a balance should be made between DC and SC generation when choosing $N_{exp}$. Also, $\smash{R_{SC, 1e^-}}$ is proportional to the number of effective transfers, $N_{trans}$; therefore, performing binning in the parallel registers decreases its contribution. For example, by doing $1\times10$ binning, the number of effective transfers per pixel is $(N_{ser}/10)+N_{par}$, instead of $N_{ser}+N_{par}$ when reading in $1\times1$ mode. Binning implies a loss in the spatial resolution, however, it decreases the readout time and, consequently, the DC contribution; plus, it increases the signal-to-noise ratio. When deciding the readout mode, an optimization of all these parameters should be made.

\begin{figure}[h!]
    \centering
    \includegraphics[width=0.5\textwidth]{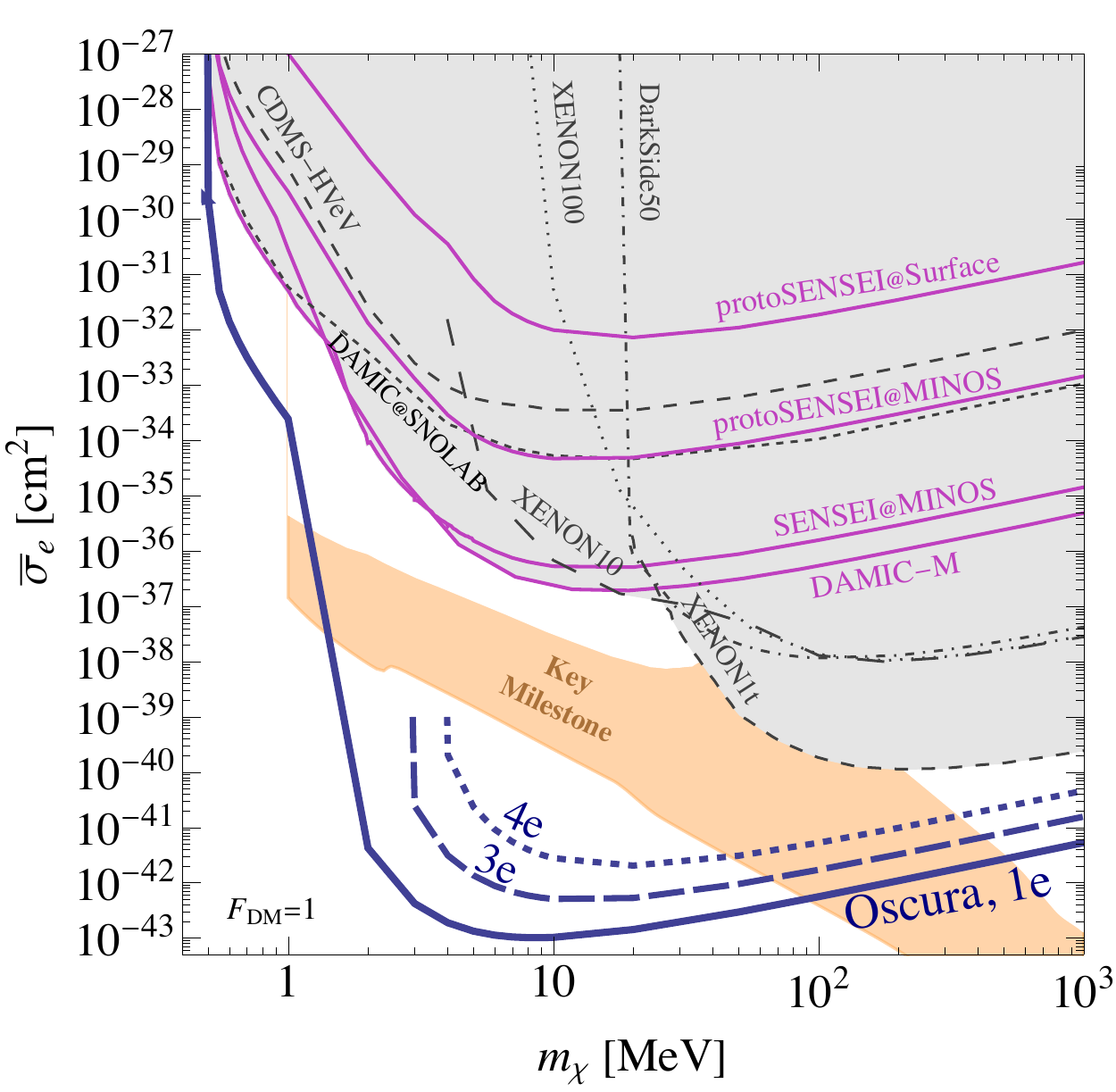}\hfill
    \includegraphics[width=0.5\textwidth]{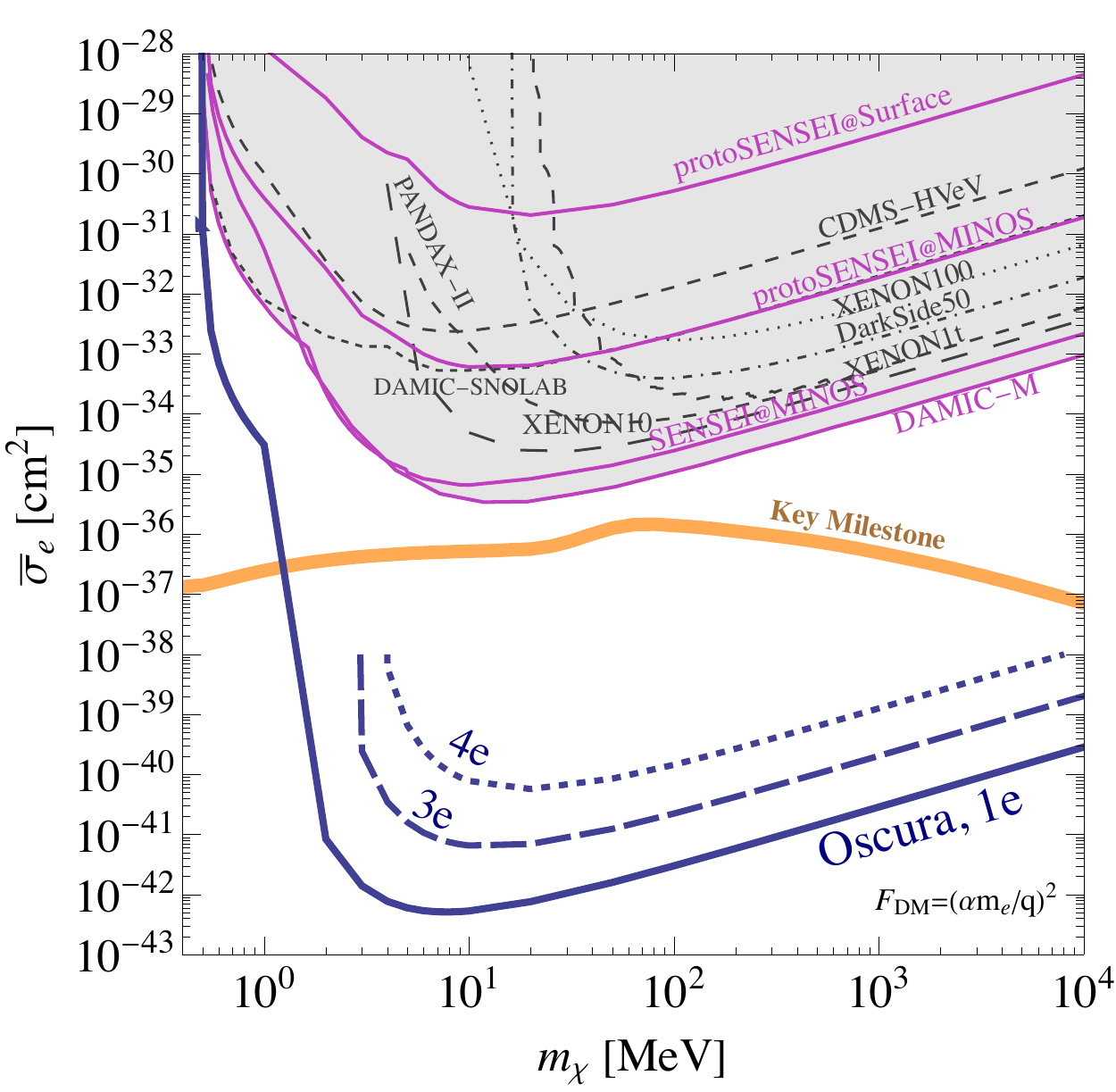}
    \caption{Approximate projected sensitivity for Oscura to DM-electron scattering at 90\% C.L. assuming a 30-kg year exposure and: 1) a $3e^-$ threshold and zero background events with $3e^-$ or more (dashed blue); 2) a $4e^-$ threshold and zero background events with $4e^-$ or more (dotted blue). To build these curves, 100\% efficiency was assumed for the reconstruction of events above the threshold. The left (right) plot assumes a heavy (light) mediator in the DM-electron interaction. The blue solid line and the other curves are as in Fig.~\ref{fig:projection-scattering}.}
    \label{fig:projection-3e}
\end{figure}

The probability of having $ne^-$ in a single pixel coming from spurious charge follows a binomial distribution. Then, the total number of $ne^-$ single pixel events from SC for the 30 kg-year Oscura exposure can be calculated as in Eq.~\ref{eq:kn}, considering $\lambda=\lambda_{SC} \equiv \kappa_{SC}\times N_{trans}$. We can estimate the Oscura instrumental background considering that we have $\smash{R_{DC, 1e^-}}=1.6\times10^{-4} e^-$/pix/day and $\kappa_{SC}=7.2\times10^{-7} e^-$/pix/transfer, consistent with the prototypes' performance. To decrease the SC contribution, we assume $N_{exp}=1$~exposure/day. In this case, we expect 0 (0) and 9.5 (0.6) events with 4$e^-$ coming from DC and SC, respectively, if performing $1\times1$ ($1\times10$) binning, for the full 30~kg-year exposure. With this level of background, the science reach of the experiment does not diminish significantly. Fig.~\ref{fig:projection-3e} illustrates Oscura science reach if unable to attain the background goal, showing the approximate projected sensitivities for Oscura considering zero background events in the 4$e^-$ bin (dotted blue line), and assuming zero events in the 3$e^-$ bin (dashed blue line).

\acknowledgments
This document was prepared by members of the Oscura collaboration using the resources of the Fermi National Accelerator Laboratory (Fermilab), a U.S. Department of Energy, Office of Science, HEP User Facility. Fermilab is managed by Fermi Research Alliance, LLC (FRA), acting under Contract No. DE-AC02-07CH11359. Also, part of this work was performed at the Center for Nanoscale Materials, a U.S. Department of Energy Office of Science User Facility, and was supported by the U.S. DOE, Office of Basic Energy Sciences, under Contract No. DE-AC02-06CH11357.

\bibliographystyle{JHEP}
\bibliography{References.bib}

\end{document}